\documentclass[preprint,showpacs,preprintnumbers,amsmath,amssymb,graphicx]{revtex4}
\usepackage{amsmath,amsfonts,amssymb}
\usepackage{epsfig}
\usepackage{color}
\def\bit{\begin{itemize}}
\def\eit{\end{itemize}}
\def\ben{\begin{enumerate}}
\def\een{\end{enumerate}}
\def\bed{\begin{description}}
\def\eed{\end{description}}

\def\k{\kappa}
\def\l{\lambda}

\def\cmg{\, {\rm cm^2/g} }

\def\half{\frac{1}{2}\,}
\def\third{\frac{1}{3}\,}

\def\lsim{\raise0.3ex\hbox{$<$\kern-0.75em\raise-1.1ex\hbox{$\sim$}}}
\def\gsim{\raise0.3ex\hbox{$>$\kern-0.75em\raise-1.1ex\hbox{$\sim$}}}

\let\jnfont=\rm
\def\NPB#1,{{\jnfont Nucl.\ Phys.\ B}{\bf #1},}
\def\PLB#1,{{\jnfont Phys.\ Lett.\ B}{\bf #1},}
\def\EPJC#1,{{\jnfont Eur.\ Phys.\ Jour.\ C}{\bf #1},}
\def\PRD#1,{{\jnfont Phys.\ Rev.\ D}{\bf #1},}
\def\PRL#1,{{\jnfont Phys.\ Rev.\ Lett.\ }{\bf #1},}
\def\MPLA#1,{{\jnfont Mod.\ Phys.\ Lett.\ A}{\bf #1},}
\def\JPG#1,{{\jnfont J.\ Phys.\ G}{\bf #1},}
\def\CTP#1,{{\jnfont Commun.\ Theor.\ Phys.\ }{\bf #1},}
\def\JHEP#1,{{\jnfont JHEP }{\bf #1},}
\def\NPPS#1,{{\jnfont Nucl.\ Phys.\ Proc.\ Suppl.\ }{\bf #1},}

\def\beq{\begin{equation}}
\def\eeq{\end{equation}}
\def\bea{\begin{eqnarray}}
\def\eea{\end{eqnarray}}
\newcommand{\ba}{\begin{array}}
\newcommand{\ea}{\end{array}}
\def\nn{\nonumber}

\begin{document}
\title{Extending the MSSM with singlet Higgs
and right handed neutrino for the self-interacting dark matter}

\author{Hai-Jing Kang, Wenyu Wang}

\affiliation{
Institute of Theoretical Physics, College of Applied Science,
     Beijing University of Technology, Beijing 100124, China}

\begin{abstract}
In order to meet the requirement of BBN, the right handed neutrino is added to
the singlet Higgs sector in the GNMSSM. The spectrum and Feynman rules
are calculated. the dark matter phenomenology is also studied. In case of $\l\sim 0$,
the singlet sector can give perfect explanation of relic abundance of dark matter
and small cosmological structure simulations. The BBN constraints on the 
light mediator can be easily solved by decaying to the right handed neutrino. 
When the $\l_N$ is at the order of $\mathcal{O}(0.1)$, the mass of the 
mediator can be constrained to several MeV.
\end{abstract}
\maketitle
\section{Introduction}
Supersymmetry (SUSY) \cite{susy,susyhierarchy} gives a natural
solution to the hierarchy problem suffered by the Standard Model
(SM). It provides a good dark matter (DM) candidate and realizes
the gauge coupling unification.  Among the SUSY models, the Minimal
Supersymmetric Standard Model (MSSM) \cite{mssm} has been
intensively studied. However, the MSSM suffers from the $\mu$-problem \cite{muew}.
In order to solve such a problem, a singlet field $S$ is always introduced to 
replace the $\mu$ term of the MSSM, the resulting model is the Next-to-Minimal
Supersymmetric Standard Model(NMSSM).\cite{fayet,NMSSM}
The effective $\mu$-term is generated by the dynamical generation of the $\mu$-term through
the coupling $S H_u H_d$ when $S$ develops a vacuum expectation value (VEV),
Nevertheless, since the Higgs and dark matter
have the singlet components,  many other problems in the MSSM
can also be relaxed. For example, the little hierarchy problem
of the MSSM is relaxed by the generation of an extra
tree-level mass term for the SM-like Higgs boson.

Since the gauge interaction of singlet dominant Higgs almost
vanishes, its mass can be tuned to any value. This is a special
feature of the NMSSM, because of that  the mass of Higgs
can be  much smaller than the electro-weak (EW) scale,
The interchange of the light Higgs  gives
a conceivably self-interaction of the singlet sector.
In a previous  paper\cite{Wang:2014kja} we studied the possibility of
the DM self-interaction to solve the small cosmological scale anomalies
in the singlet extensions of the MSSM.\cite{Bringmann:2013vra}
We found that the correlation between the DM
annihilation rate and DM-nucleon spin independent (SI) cross section
strongly constrains this model so that it cannot realize the DM self-interacting
scenario in the NMSSM. After that, we found that DM self-interacting
scenario can be realized by extending the singlet  most generally (denoted as GNMSSM).
In case of $\lambda\sim 0$, the singlet forms a
dark sector. DM is singlino, light CP-even Higgs is the mediator.
Enough large parameter space survives and the Sommerfeld enhancement factor
can be realized too.

As pointed in \cite{Kaplinghat:2013yxa},  the singlet CP-even Higgs can not be exactly dark,
it must decay before the start of the  Big Bang Nucleosynthesis (BBN) ($\sim 1$ sec)
so the decay products will not affect BBN.
This constraint sets a minimum interaction coupling between the Standard Model
and the dark sector in order to facilitate the fast decay of the singlet Higgs
before the BBN era. But if the singlet Higgs couples to the
SM particles, the direct detection rate of DM  will be enhanced greatly
by the light mediator,\cite{Wang:2014kja} thus it will be excluded by 
measurement of  the SI cross section. This seems to be an obstacle of the model.

One way to solve such an obstacle is to add a right handed
neutrino to the GNMSSM, in which the singlet Higgs decay into
right handed sterile neutrino, and it has nothing to do with
the quarks, escaping from the detection rate.
In fact, there should be such a singlet in the particle spectrum
since the oscillation of neutrino implies that mass and right handed neutrino
probably exists. A realistic SUSY model must have the massive neutrinos, too.
The most economical introduction should be replacing
the singlet Higgs with singlet right-handed neutrino superfield ($\widehat N$).\cite{mnmssm}
However, such a model violates the R-parity, giving no DM candidate.
In order to preserve R-parity, $\widehat N$ must be an additional supermultiplet.
A TeV-scale Majorana mass for the right-handed neutrino 
is dynamically generated through the $\widehat{S}\widehat{N}\widehat{N}$
coupling when $S$ develops a VEV (note that such a TeV-scale
majorana mass is too low for the see-saw mechanism and thus the
neutrino Yukawa couplings $H_u L N$ must be very small). In the same
way, a TeV-scale mass for the right-handed sneutrino can also be
generated, which can serve as a good dark matter candidate
\cite{Cerdeno:2009dv}. In paper \cite{Wang:2013jya}, we find the right handed
neutrino can enhance the mass of SM Higgs several GeV.

In this work we will study the SUSY model of  singlet Higgs and right handed neutrino, and check
if the self-interacting DM scenario and BBN constraints
can be realized. We will take into account of constraints from DM relic abundance.
We organize the content as follows. In Sec. \ref{sec2}, we will show the details of the model
and discuss the general interactions of DM. In Sec. \ref{sec3} we show
the numerical results, and conclusion is given in  Sec. \ref{sec4}.

\section{Model, spectrum and Feynman rules}\label{sec2}
As talked in the introduction, the $\mu$ term can not be replaced
by singlet right handed neutrino $\widehat{N}$ in the GNMSSM,
and there must be a singlet Higgs $\widehat{S}$. The superpotential is \cite{Wang:2009rj}
\bea
W = W_{\rm Yukawa} +\eta \widehat{S}
+\half \mu_s \widehat{S}^2 + \frac{1}{3} \k \widehat{S}^3 + \half \mu_N\widehat{N}^2
 + \l_N \widehat{S}\widehat{N}^2\ ,\label{grn-sp}
\eea
where
\begin{eqnarray}
 W_{\rm Yukawa} = Y_u \widehat{H}_u \cdot \widehat{Q} \widehat{u}_R
- Y_d \widehat{H}_d \cdot \widehat{Q} \widehat{d}_R   - Y_e \widehat{H}_d \cdot \widehat{L} \widehat{e}_R
+ Y_\nu \widehat{H}_u \cdot \widehat{L} \widehat{N}
   + \lambda \widehat{S} \widehat{H}_u \cdot \widehat{H}_d.  \label{sp-nm}
\end{eqnarray}
Note that here we impose R-parity and thus the linear and triple terms of  $\widehat{N}$
are forbidden. As a result, there is no VEV for the right-handed sneutrino
(we will show how to get the global minimum in the following).
We set $\l\sim 0$ so that the singlet sector decouples from the SM sector.
In the following discussion, we will concentrate on the singlet sector.
The soft SUSY breaking terms take the form
 \bea
 -{\cal L}_\mathrm{soft} &=&  m_s^2 | S |^2
 + \left( C_\eta \eta S +\half B_s \mu_s S^2 + \third \k A_\kappa S^3 + \mathrm{h.c.}
 \right)\label{soft-gm}\\
&+&  m_{\tilde N}^2 | {\tilde N} |^2  +\left(\half B_N \mu_N {\tilde N}^2
 + \l_N A_N S{\tilde N}^2+\mathrm{h.c.}\right)\,.
\nn\eea
The first line is the  soft terms  of the singlet Higgs, while the second
line is of the right handed sneutrino.
The main difference between the NMSSM and GNMSSM is reflected in their Higgs sectors
which contain different singlet Higgs mass matrices and self-interactions.
The difference mainly comes from the F-term potential:
\bea
V_{F_{\rm S}} &=&  |\eta+\mu_s S +\k S^2+\lambda_N {\tilde N}^2|^2\nn\\
&=& |\k S^2|^2 + \eta^2+\mu_s^2|S|^2
+\left(\eta\mu_s S  +\k \eta S^2 +\k\mu_s S^2 S^\ast + \mathrm{h.c.}\right)\\
V_{F_{\rm N}} &=&  |2\l_N S+\mu_N|^2 |{\tilde N}|^2 .
\eea
Since $\eta^2$ is a constant, the $\mu_s^2|S|^2, \eta\mu_s S,\k \eta S^2$ terms
can be absorbed by the redefinition of the soft SUSY breaking parameters
$m_s^2|S|^2,~ C_\eta \eta S,\half B_s \mu_s S^2$. Then,
the singlet potential is
\bea
V & = & V_F+V_{\rm soft}\nn\\
  & = &   m_s^2|S|^2 + \left(C_\eta \eta S+\half B_s \mu_sS^2
+ \third \k A_\k\ S^3 +\k\mu_s S^2 S^\ast + \mathrm{h.c.}\right)\nn\\
& + & |2\l_N S+\mu_N|^2 |{\tilde N}|^2+m_N^2 | {\tilde N} |^2  +\left(\half B_N \mu_N {\tilde N}^2
 + \l_N A_N S{\tilde N}^2+\mathrm{h.c.}\right)
\label{vform}
\eea

The spectrum of the model is simple, The chiral supermultiplet of singlet Higgs
contains a complex scalar and a Majorana fermion $\chi$. After the scalar
component getting a VEV $v_s$,  we can get one CP-even Higgs $h$ and
one CP-odd Higgs $a$. The mass spectrum and the relevant Feynman rules are
\bea
m_{\chi}&=&2\k v_s+\mu_{S}, \\
m_{h}^{2}&=&\k v_ss(4\k v_s+A_{\k}+3\mu_{S})-C_\eta\eta /v_s, \\
m_{a}^{2}&=&-2B_{s}\mu_{s}-\k v_s(3A_{\k}+\mu_{S})-C_\eta\eta/v_s,
\label{gm-fr}
\eea
As for the neutrino sector, a  Majorana mass can be 
generated through the coupling $SN^2$ and $\mu_N N^2 $. In this paper
we set $\mu_N$ at TeV scale, thus the  Majorana
mass is too small for the conventional see-saw mechanism and
the Yukawa coupling $y_\nu H_u L N$ should be very small ($y_N\ll 1$)
and be neglected. Since there is no Dirac mass term here, the mass spectrum
of the right-handed neutrino sector is also very simple.
Denoting $\tilde N=R+iM$, there are only one
CP-even right-handed sneutrino (denoted as $R$)
and one CP-odd right-handed sneutrino (denoted as $M$).
The right-handed neutrino is denoted as $N$.
From Eq. (\ref{vform}), we can get the spectrum as
\bea
m_R^2 &=& (2\l_N v_s+\mu_N)^2 + M_{\tilde N}^2 + 2\l_N v_s A_N
+ 2\l_N(\k v_s^2 - \l v_u v_d)\nn\\
m_M^2 &=& (2\l_N v_s+\mu_N)^2 + M_{\tilde N}^2 - 2\l_N v_s A_N
- 2\l_N(\k v_s^2 - \l v_u v_d) \nn\\
m_N &=& \mu_N +2\l_N v_s. \label{rnspc}
\eea
Note that in our numerical study we require $M_R^2$ and $M_M^2$
be positive, and, as a result, the global minimum of the scalar
potential locates at the zero point of the right-handed
sneutrino field (the right-handed sneutrino has no VEV and R-parity is preserved).

With the above spectrum we can get the couplings between
the Higgs and the right-handed neutrino/sneutrino. We list
the corresponding Feynman Rules for the following calculation:
\bea
V_{hhh} &=& -\sqrt{2}k(6\k s+A_{\k}+3\mu_{S})=-\sqrt{2}\k(3m_{\chi}+A_{\k}),\\
V_{haa} &=& -\sqrt{2}\k(2\k s-A_{\k}+\mu_{S})=-\sqrt{2}\k(m_{\chi}-A_{\k}),\\
V_{h\chi\chi}&=&-\sqrt{2}\k,\\
V_{a\chi\chi}&=&\sqrt{2}i\k\gamma^{5},
\eea
\bea
V_{h N N}&=&-\sqrt{2}\lambda_{N},\\
V_{a N N}&=&\sqrt{2}i\lambda_{N}\gamma^{5},\\
V_{R\chi N}&=&-\frac{1}{\sqrt{2}}\lambda_{N},\\
V_{M\chi N}&=&\frac{i}{\sqrt{2}}\lambda_{N}\gamma^{5}.
\eea
Note that lighter right handed sneutino can be a DM candidate.
In this work, we require that the LSP is singlino $\chi$.
As talked in the introduction,  the predictions of the $\Lambda$CDM model
on small cosmological scale structures seem not so successful and there are mainly three anomalies:
 \textit{missing satellites}, \textit{cusp vs ~core}, \textit{too big to fail}.
\cite{Klypin:1999uc,deNaray:2011hy,BoylanKolchin:2011de}.
To solve these problems, it needs a proper self-interaction between dark matter
that can give the non-relativistic self-scattering cross section.\cite{Tulin:2012wi}
The scattering cross section between DM can be described by quantum mechanics.
In our model, interchanging $h$ between  singlino $\chi$ forms a Yukawa potential
\bea
V(r)=\frac{\kappa^2}{2\pi}e^{-m_h r}.\label{Vyukawa}
\eea
Since mass of $h$ can be tuned to any value,
self interaction can be realized  without right handed neutrino. we will check
the case of participation of right handed neutrino.
Also we will set $N$ lighter than $m_h/2$ so that  $h$ can decay
to the right handed neutrino $N$ to satisfy the BBN constraint. \cite{Kouvaris:2014uoa}
The numerical results are shown in the following section.
\section{Realizing the self-interacting dark matter}
\label{sec3}
The numerical input for the simulation of small scales is the differential cross section
$d \sigma/ d \Omega$ as a function of the DM relative velocity $v$.
The standard cross section $\sigma = \int d\Omega (d\sigma/d\Omega)$ receives a strong
enhancement in the forward backward scattering
limit ($\cos\theta \to \pm 1$), which does not change the DM particle trajectories.
Thus two additional cross sections are defined to parameterize
transport~\cite{krstic:1999}, the transfer cross section $\sigma_T$ and the viscosity
 (or conductivity) cross section $\sigma_V$:\cite{Tulin:2012wi,Tulin:2013teo,Ko:2014nha}
\beq
\sigma_T =  \int d\Omega \, (1-\cos\theta) \,\frac{d\sigma}{d\Omega}  \, , \qquad  \label{sigmaT}
\sigma_V =  \int d \Omega \, \sin^2 \theta \, \frac{d\sigma}{d\Omega} \, .
\eeq
$\sigma_T$ is for the estimation of the Dirac DM, while $\sigma_T$ is for the Majorana DM.
Since the singlino $\chi$ is a Majorana fermion, thus we should check the
 viscosity cross section which should be defined with two variables:
\bea
\frac{d\sigma_{VS}}{d\Omega} &=& \left|f(\theta)+f(\pi-\theta)\right|^2 = \frac{1}{k^2}\left|
\sum_{\ell (\mbox{\tiny\rm EVEN\  number})}^{\infty} (2\ell + 1) (\exp(2i\delta_l)-1)P_\ell(\cos\theta)\right|^2
 \label{sigmaVSsum}\\
\frac{d\sigma_{VA}}{d\Omega} &=& \left|f(\theta)-f(\pi-\theta)\right|^2 = \frac{1}{k^2}\left|
\sum_{\ell (\mbox{\tiny\rm ODD\ number})}^{\infty} (2\ell + 1) (\exp(2i\delta_l)-1)P_\ell(\cos\theta)\right|^2
\label{sigmaVAsum}
\eea
Using the orthogonality relation for the Legendre polynomials, we can obtain
\bea
\frac{\sigma_{VS}k^2}{4\pi} &=& \sum_{\ell(\mbox{\tiny EVEN\ number})}^{\infty}4\sin^{2}(\delta_{\ell+2}
-\delta_{\ell})(\ell+1)(\ell+2)/(2\ell+3) ,\label{sigmaVS}\\
\frac{\sigma_{VA}k^2}{4\pi} &=& \sum_{\ell(\mbox{\tiny ODD\ number})}^{\infty}4\sin^{2}(\delta_{\ell+2}
-\delta_{\ell})(\ell+1)(\ell+2)/(2\ell+3) . \label{sigmaVA1}
\eea
The phase shift $\delta_{\ell}$ must be computed by solving the Schr\"{o}dinger equation
$$
\frac{1}{r^2} \frac{d}{dr} \Big( r^2 \frac{d R_{\ell}}{dr} \Big)
+ \Big( k^2 - \frac{\ell (\ell + 1)}{r^2} - 2m_r V(r) \Big) R_\ell = 0
$$
directly with partial  wave expansion method.
The total wave function of the  spin-1/2 fermionic DM must
be antisymmetric with respect to the interchange of two
identical particles. Then the spatial wave function should
be symmetric when the total spin is 0 (singlet)
while the spatial wave function should be antisymmetric when the total spin is 1 (triplet).
In our following analysis, we assume that the DM scatters
with random orientations, thus the triplet is three times as likely as the
singlet and the average cross section will be
\bea
\sigma_{V} =\frac{1}{4}\sigma_{VS}+\frac{3}{4} \sigma_{VA}\; .\label{sigmaVa;;}
\eea

\begin{figure}
\begin{center}\scalebox{1}{\epsfig{file=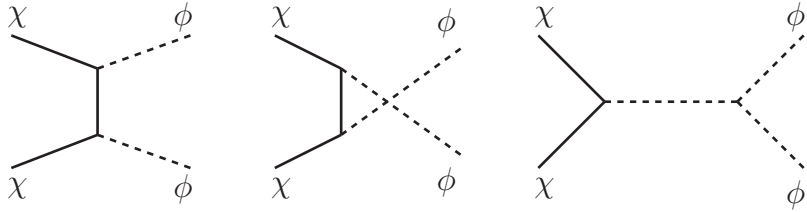}}\end{center}
\vspace{-0.7cm}
\caption{Diagram of dark matter annihilation to singlet Higgs $h, a$.}
\label{fig1}
\end{figure}

\begin{figure}
\begin{center}\scalebox{1}{\epsfig{file=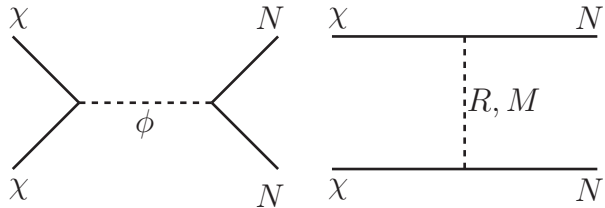}}\end{center}
\vspace{-0.7cm}
\caption{Diagram of dark matter annihilation to right handed neutrino $N$.}
\label{fig2}
\end{figure}

Since $\lambda\sim 0$ we do not need to take care of  the direct detection rate of DM,
but the observed relic abundence of DM  must be considered. In our model,
$\chi$ can annihilate to  singlet Higgs and right handed neutrino, and the rate
determines the relic density which gives a stringent constraints on the model.
The general form of the annihilation cross section is given by \cite{Drees:1992am, susy-dm-review}
\beq
\sigma v =\frac{1}{4}\frac{\bar\beta_f}{8\pi sS}\left[
|A(^1S_0)|^2+\frac{1}{3}\left(|A(^3P_0)|^2+|A(^3P_1)|^2\right)+|A(^3P_2)|^2\right],
\eeq
where $S$ is the symmetry factor, $A(^1S_0),~A(^3P_0),~A(^3P_1),~A(^3P_2)$
are the contributions from different spin states of DM and $\beta_f$ is given by
\bea
\bar\beta_f=\sqrt{1-2(m_X^2+m_Y^2)/s + (m_X^2+m_Y^2)^2/s^2}~,
\eea
with $X,Y$ being the final states. The Feynman diagrams of the annihilation are shown
in  Fig.\ref{fig1} and Fig.\ref{fig2}.  Fig.\ref{fig1} shows the annihilation
to the singlet Higgs, in which $\phi$ can be $h$ or $a$ or both.
 Fig.\ref{fig2} shows the annihilation to the right handed neutrino $N$.
The analytical results of $\sigma v$ is shown in the appendix.

\begin{figure}
\begin{center}
\scalebox{1.3}{\epsfig{file=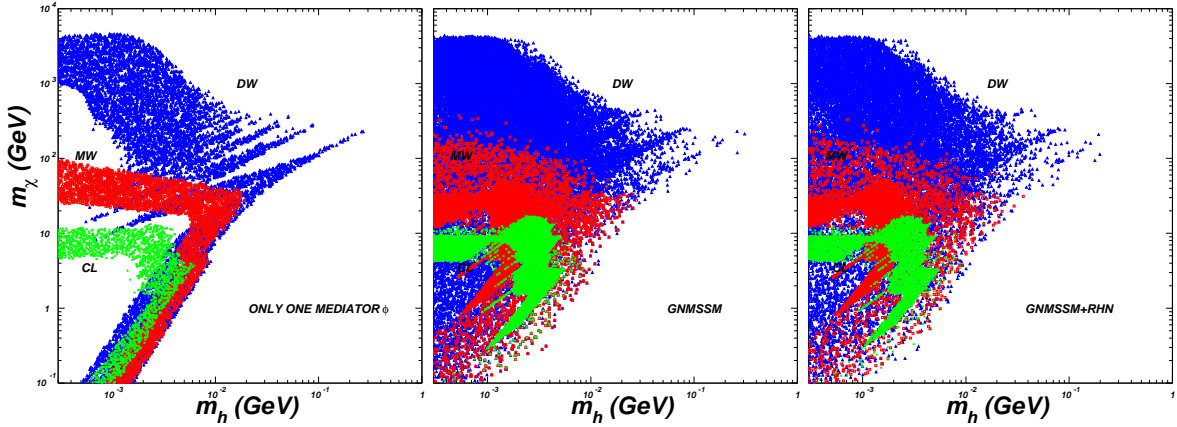}}
\end{center}
\caption{The survived points under the  contraint of relic density,
scattering cross section and BBN.
The blue points are the points satisfying the simulation in the Dwarf scale
($\sigma/m_\chi \sim 0.1 - 10 \, \cmg$, the characteristic velocity is 10 km/s.)
The red points are the points satisfying the simulation in the Milky Way
($\sigma/m_\chi \sim 0.1 - 1 \, \cmg$, the characteristic velocity is 200 km/s.)
The green points are the points satisfying the simulation in the Milky Way
($\sigma/m_\chi \sim 0.1 - 1 \, \cmg$, the characteristic velocity is 1000 km/s.)}
\label{fig3}
\end{figure}

Next we do the numerical calculation to see if the model can survive under the
constraints of relic density, scattering cross section and BBN.
From the spectrum and Feynman rules above, we can see that the
mass parameters of the six particles ($h, a,  \chi, R, M, N$)
can be set freely  because the relevant input soft parameters can take arbitrary values.
This makes the following calculation much easier and we can choose
the input parameters easily without fine-tuning.
Then we have  nine input parameters in our calculation,
namely
\beq
 m_\chi,~m_h,~m_a,~m_R,~m_M,~m_N,~A_\k,~\k,~\l_N.
\eeq
The first seven parameters are mass dimension, we scan them in the
range of $(10^{-4},~10^{4})$ GeV, the last two parameters are dimensionless,
we scan them in the range of $(10^{-11},~10^{1})$. Note that
getting a mass much lighter than EW scale generally needs some tuning.
In our scan we set $m_R, m_M > m_\chi$  and $m_N < m_h/2$.
The numberical results are shown in Fig.\ref{fig3},
which shows the  survived points in $(m_\chi,~m_h)$ space.
The left plot shows the  case of only one mediator scalar $h$,
the middle plot shows the case of only singlet Higgs in the GNMSSM,
while the right plots shows the results of the model above which
is GNMSSM plus a right handed neutrino.

From Fig.\ref{fig3}, we can see that the simulation of small structure gives
a stringent constraints on the self interacting DM, especially, when there is only
one light mediator. However, there still are points left which satisfy all the requirements
of Dwarf scale, Milky Way and cluster. GNMSSM without  right handed neutrino
has a larger parameter space to solve the anomalies of  all three small scales.
The reason is that in the DM self-interaction model \cite{Tulin:2013teo}
DM can only annihilate into $hh$ via $t$-channel and $u$-channel while in the GNMSSM
DM can annihilate into $hh$, $ha$ and $aa$ via $t$-channel, $u$-channel and $s$-channel,
as shown in Fig.\ref{fig1}. The results of GNMSSM with right handed neutrino are almost
the same as that without it. The reason is that the BBN constraints in fact give a
much lower bound on the coupling strength $\l_N$. The width of $h$ in this
case is
\bea
\Gamma_h &=& \l_N^2\frac{\sqrt{m_{h}^{2}/4-m_N^{2}}}{2\pi m_{h}^{2}}(m_{h}^{2}-4m_N^{2})\\
&\simeq& \l_N^2\frac{m_h}{4\pi},~~~~~~({\rm in~case~of~} m_h\gg m_N).
\eea
One can check that if $\l_N$ is greater than $10^{-10}$, the mediator (10 MeV) can decay much earlier
before 1 second, thus it is much easy to meet the BBN requirement.

\begin{figure}
\begin{center}
\scalebox{0.35}{\epsfig{file=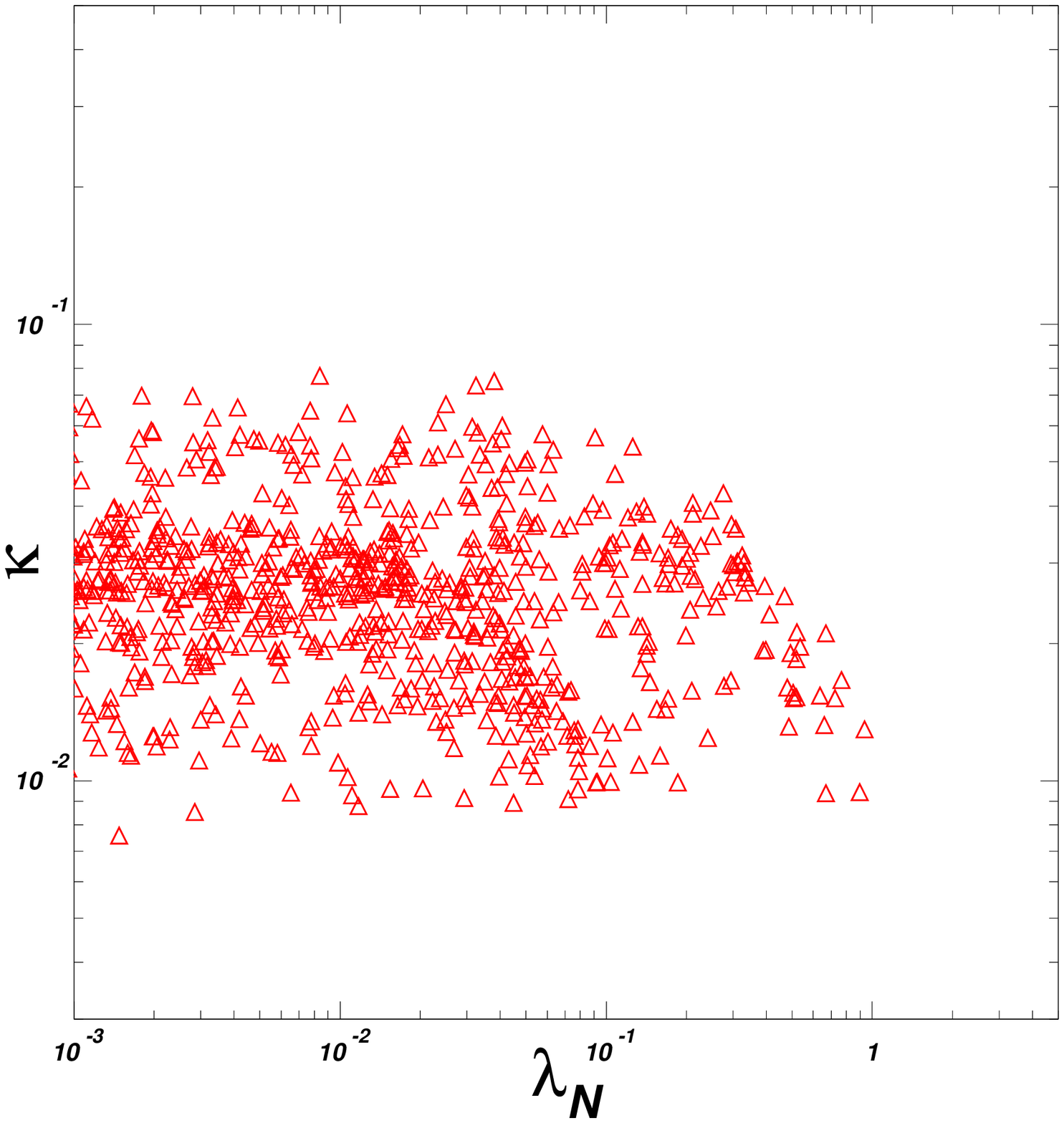}}
\scalebox{0.35}{\epsfig{file=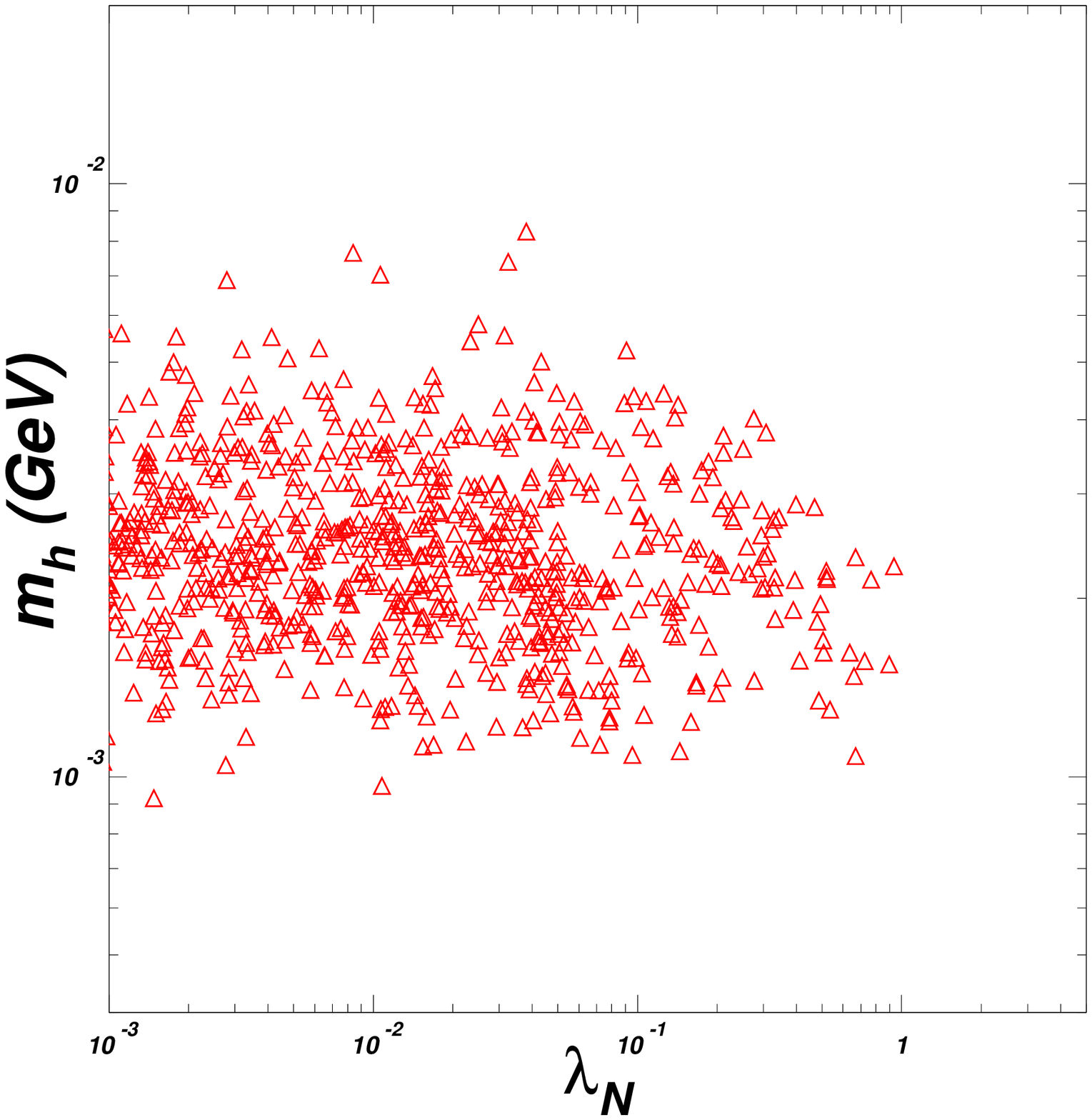}}
\end{center}
\vspace{-0.7cm}
\caption{Parameters $m_h$, $\k$ versus $\l_N$ of the survived points.}
\label{fig4}
\end{figure}
Though BBN constraints on $\l_N$ is loose, the survived parameter
space of the model can be changed when $\l_N$ is at order of $\mathcal{O}(0.1)$
 the annihilation to $N$ become significant. This can been seen from the left plot
of Fig. \ref{fig4} that when $\l_N$ increases up to $\mathcal{O}(0.1)$, $\k$ begins to decrease
to a small value. In this case, the annihilation channels to $N$ (shown in Fig.\ref{fig2})
give proper contribution to the relic density of DM.
The right plot shows survived points in $(m_h,~\l_N)$ space. We can see that
as $\l_N$ increases, the mass of mediator $h$ is constraint to several MeV.
Moreover an interaction strength $\l_{N} \sim \mathcal{O}(0.1)$ has been advocated
as a way of suppressing the standard active-to-sterile oscillation production process,
easing the cosmological constraints from $N_{{\rm eff}}$.\cite{Dasgupta:2013zpn}
Note that the right handed sterile neutrino  $N$  must decay prior to BBN too.
The dominant decay channel is to three active neutrinos with rate
\begin{eqnarray}
\Gamma_{3 \nu} \simeq \frac{G_{F}^{2} m_{N}^{5}}{192 \pi^{3}} \sin^{2} \Theta,
\end{eqnarray}
where $\sin \Theta$ is the sterile-active mixing angle. A proper 
$\sin \Theta$ ($>10^{-3}$) can make sure that a MeV order $N$ 
decays at a pre-BBN time.\cite{Kusenko:2009up} In all, we conclude that
adding right handed neutrino to the GNMSSM gives  
a satisfactory solution to the otherwise problematic decay of $h$.
\section{Conclusions}\label{sec4}
In this work, in order to solve the BBN constraints on the singlet Higgs
we add the right handed neutrino to the GNMSSM. we calculated the
spectrum and Feynman rules. and studied the the DM phenomenalogy.
In case the $\l\sim 0$ the singlet sector can give perfect explanation
of relic abundance and small structure simulations. The BBN constraints
 on the light mediator $h$ can be easily solved by decaying
to the right handed neutrino.  In case of the $\l_N$ at the order of $\mathcal{O}(0.1)$, 
the mass of the mediator mass can be constrained to several MeV. 
Since we set $\l \sim 0$ the direct detection rate doesn't need to be considered.
In case of  $\l \ne 0$, the singlet Higgs and right handed neutrino will interact
with the gauge sector, the resusts will appear in our further studies.
Note that recently both ATLAS and CMS reported a hint for a diphoton resonance ($X$) 
around 750 GeV  with width about 40GeV.\cite{atlas-diphoton,cms-diphoton}
Such a resonance can be explained in many SUSY models,\cite{diphoton-susy} among which 
$X$ particle can be interpreted as the heavy Higgs in the NMSSM.
Since right handed neutrino is a gauge singlet, it will increase the 
the invisible decay width in our model. Detailed study of the phenomenology
is beyond this work.

\section*{Acknowledgments}
This work was supported by the
Natural Science Foundation of China under grant numbers 11375001
and Ri-Xin Foundation of BJUT by talents foundation of education department of Beijing.
\section*{Appendix}
The amplitudes from different final states are given by
\begin{enumerate}
\item $\chi\chi\to hh$ :
\bea
A(^3P_0) &=& 2\sqrt{6}v\k^2\left[\frac{R(3m_\chi+A_\k)}{4-R(m_h)^2+iG_h}
-2\frac{1+R(m_\chi)}{P_\chi}+\frac{4}{3}\frac{\bar\beta_f^2}{P_\chi^2}\right] ,\\
A(^3P_2) &=& -(16/\sqrt{3})v\k^2\bar\beta_f^2 /P_\chi^2~.
\eea
\item   $\chi\chi\to aa$ :
\bea
A(^3P_0) &=& 2\sqrt{6}v\k^2\left[\frac{R(m_\chi-A_\k)}{4-R(m_h)^2+iG_h}
-2\frac{1-R(m_\chi)}{P_\chi}+\frac{4}{3}\frac{\bar\beta_f^2}{P_\chi^2}\right], \\
A(^3P_2) &=& -(16/\sqrt{3}) v\k^2\bar\beta_f^2 /P_\chi^2~.
\eea
\item   $\chi\chi\to ha$ :
\bea
A(^1S_0) &=& -4\sqrt{2}\k^2\frac{R(m_\chi-A_\k)}{4-R(m_a)^2}\left(1+\frac{v^2}{8}\right)
\nn\\
&& +8\sqrt{2}\k^2\frac{R(m_\chi)}{P_\chi}\left[1+v^2\left(\frac{1}{8}
-\frac{1}{2P_\chi}+\frac{\bar\beta_f^2}{3P_\chi^2}\right)\right]\nn\\
&& +2\sqrt{2}\k^2 \left(R(m_a)^2-R(m_h)^2\right)\left[1+v^2\left(-\frac{1}{8}
-\frac{1}{2P_\chi}+\frac{\bar\beta_f^2}{3P_\chi^2}\right)\right], \\
A(^3P_1) &=& 8v\k^2 \bar\beta_f^2 /P_\chi^2~.
\eea
\item   $\chi\chi\to NN$:
\bea
A(^1S_0) &=& \lambda_N^2 \left[1+v^2\left(-\frac{1}{2P_R}
+\frac{\bar\beta_f^2}{3P_R^2}\right)\right]R(m_N)/P_R\nn\\
&+& \lambda_N^2 \frac{1}{P_R}\left[1+v^2\left(\frac{1}{4}-\frac{1}{2P_R}
-\frac{\bar\beta_f^2}{6P_R}+\frac{\bar\beta_f^2}{3P_R^2}\right)\right]\nn\\
&-& \lambda_N^2 \left[1+v^2\left(-\frac{1}{2P_M}
+\frac{\bar\beta_f^2}{3P_M^2}\right)\right]R(m_N)/P_M\nn\\
&+& \lambda_N^2 \frac{1}{P_M}\left[1+v^2\left(\frac{1}{4}-\frac{1}{2P_M}
-\frac{\bar\beta_f^2}{6P_M}+\frac{\bar\beta_f^2}{3P_M^2}\right)\right]\nn\\
&-& 8\kappa\lambda_N\frac{1}{4-R(m_a)^2+iG_a}\left(1+\frac{v^2}{4}\right).
\eea
\bea
A(^3P_{00}) &=& \l_N^2\bar\beta_f\left[-\frac{1}{2}\left(\frac{1}{2P_R}
-\frac{2}{3P_R^2}\right) + \frac{R(m_N)}{P_R^2}-\frac{1}{2}\left(\frac{1}{2P_M}
-\frac{2}{3P_M^2}\right) - \frac{R(m_N)}{P_M^2}\right] \nn\\
&-& 2\kappa\lambda_N\bar\beta_f\frac{1}{4-R(m_h)^2+iG_h}
+ 2\kappa\lambda_N\bar\beta_f\frac{1}{4-R(m_a)^2+iG_a}
\eea
\bea
A(^3P_{10}) &=& 0.
\eea
\bea
A(^3P_{11})(\l') &=& \l_N^2 \l'\bar\beta_f\left(-\frac{1}{P_R}+\frac{1}{P_R^2}+\frac{R(m_N)}{P_R^2}\right)
- \l_N^2 \l'\bar\beta_f\left(-\frac{1}{P_M}+\frac{1}{P_M^2}-\frac{R(m_N)}{P_M^2}\right).\nn\\
\eea
\bea
A(^3P_{20}) &=& \l_N^2\bar\beta_f \left(\frac{R(m_N)+1}{P_R^2}-\frac{R(m_N)-1}{P_M^2}\right).
\eea
\bea
A(^3P_{21}) &=& \l_N^2\bar\beta_f\left(-\frac{1+R(m_N)}{P_R^2}+\frac{1-R(m_N)}{P_M^2}\right).
\eea
\end{enumerate}
In the above formulas
\bea
R(m_X)=\frac{m_X}{m_\chi},~P_j=1+R(m_j)^2-\frac{1}{2}(R(m_X)^2+R(m_Y)^2),
~G_i=\frac{\Gamma_i m_i}{m_\chi^2}
\eea
and  $\l'$ is the number of the final helicity state.


\begin{thebibliography}{}
\bibitem{susy}
J.~Wess and B.~Zumino, Nucl. Phys. {\bf B70} (1974) 39;
M.F.~Sohnius, Phys. Rep. {\bf 128} (1985) 39.

\bibitem{susyhierarchy}
S. Dimopoulos and H. Georgi, Nucl. Phys. {\bf B193}  (1981) 150;
 N. Sakai, Z. Phys. {\bf C11} (1981) 153.

\bibitem{mssm} For a review, see, e.g.,
  H.~E.~Haber and G.~L.~Kane, Phys.\ Rept.\  {\bf 117}, 75 (1985).

\bibitem{muew}
   J.~E.~Kim and H.~P.~Nilles,
   Phys.\ Lett.\ B {\bf 138}, 150 (1984).

\bibitem{fayet}
   P.~Fayet,
   Nucl.\ Phys.\ B {\bf 90}, 104 (1975);
   Phys.\ Lett.\ B {\bf 69}, 489 (1977).

\bibitem{NMSSM} For pheno studies of NMSSM, see, e.g.,
   J.~R.~Ellis,  {\it{et al.}}, \PRD39, 844 (1989);
   M.~Drees, Int.\ J.\ Mod.\ Phys.\ A {\bf 4}, 3635 (1989);
   P. N. Pandita, \PLB318, 338 (1993);\PRD 50, 571 (1994);
   S.~F.~King, P.~L.~White, \PRD52, 4183 (1995);
   B. Ananthanarayan, P.N. Pandita, \PLB353, 70 (1995); \PLB371, 245 (1996);
                                    Int. J. Mod. Phys. A12, 2321 (1997);
   B. A. Dobrescu, K. T. Matchev, \JHEP0009, 031 (2000);
   V. Barger, P. Langacker, H.-S. Lee, G. Shaughnessy, \PRD73,(2006) 115010;
    R.~Dermisek, J.~F.~Gunion, \PRL95, 041801 (2005);
    G.~Hiller, \PRD70, 034018 (2004);
    F.~Domingo, U.~Ellwanger, \JHEP0712, 090 (2007);
    Z.~Heng,  {\it et al.},
    Phys.\ Rev.\ D {\bf 77}, 095012 (2008);
    R. N. Hodgkinson, A. Pilaftsis, \PRD76, 015007 (2007); \PRD78, 075004 (2008);
    W. Wang, Z. Xiong, J. M. Yang,
    Phys.\ Lett.\ B {\bf 680}, 167 (2009);
    J. Cao, J. M. Yang,
    Phys.\ Rev.\ D {\bf 78}, 115001 (2008);
    JHEP {\bf 0812}, 006 (2008);
    U.~Ellwanger, C.~Hugonie and A.~M.~Teixeira,
    Phys.\ Rept.\  {\bf 496}, 1 (2010);
    M.~Maniatis,
    Int.\ J.\ Mod.\ Phys.\ {\bf A25} (2010) 3505;
    U.~Ellwanger,
    Eur.\ Phys.\ J.\  C {\bf 71}, 1782 (2011);
    J.~Cao,  {\it et al.},
    JHEP {\bf 1311}, 018 (2013);
    JHEP {\bf 1304}, 134 (2013);
    JHEP {\bf 1309}, 043 (2013);
    JHEP {\bf 1206}, 145 (2012);
    JHEP {\bf 1210}, 079 (2012);
    JHEP {\bf 1203}, 086 (2012);
    Phys.\ Lett.\ B {\bf 703}, 462 (2011);
    JHEP {\bf 1011}, 110 (2010);
    C.~Han, {\it et al.},
    JHEP {\bf 1404}, 003 (2014);
    Z.~Kang,  {\it et al.},
    arXiv:1102.5644 [hep-ph];
    JCAP {\bf 1101}, 028 (2011);
  J.~Kozaczuk and S.~Profumo,
  arXiv:1308.5705 [hep-ph].
   Phys.\ Rev.\ D {\bf 82}, 051701 (2010). 
   J.~Cao, {\it et al.},
  arXiv:1311.0678 [hep-ph];
  Phys.\ Lett.\ B {\bf 703}, 292 (2011);
  JHEP {\bf 1007}, 044 (2010).

\bibitem{Wang:2014kja}
  F.~Wang, W.~Wang, J.~M.~Yang and S.~Zhou,
  Phys.\ Rev.\ D {\bf 90}, no. 3, 035028 (2014)
  [arXiv:1404.6705 [hep-ph]].

\bibitem{Bringmann:2013vra}
  T.~Bringmann, J.~Hasenkamp and J.~Kersten,
  arXiv:1312.4947 [hep-ph].

\bibitem{Kaplinghat:2013yxa}
  M.~Kaplinghat, S.~Tulin and H.~B.~Yu,
  Phys.\ Rev.\ D {\bf 89}, no. 3, 035009 (2014)
  [arXiv:1310.7945 [hep-ph]].

\bibitem{mnmssm}
  D.~E.~Lopez-Fogliani and C.~Munoz,
  Phys.\ Rev.\ Lett.\  {\bf 97}, 041801 (2006)
  [hep-ph/0508297];
  A.~Abada, G.~Bhattacharyya and G.~Moreau,
  Phys.\ Lett.\ B {\bf 642}, 503 (2006)
  [hep-ph/0606179];
  A.~Abada and G.~Moreau,
  JHEP {\bf 0608}, 044 (2006)
  [hep-ph/0604216].


\bibitem{Cerdeno:2009dv}
  D.~G.~Cerdeno and O.~Seto, JCAP {\bf 0908}, 032 (2009).

\bibitem{Wang:2013jya}
  W.~Wang, J.~M.~Yang and L.~L.~You,
  JHEP {\bf 1307}, 158 (2013)
  [arXiv:1303.6465 [hep-ph]].

\bibitem{Wang:2009rj}
   W.~Wang, Z.~Xiong, J.~M.~Yang and L.~-X.~Yu,
   JHEP {\bf 0911}, 053 (2009). 

\bibitem{Klypin:1999uc}
  A.~A.~Klypin, A.~V.~Kravtsov, O.~Valenzuela and F.~Prada,
  Astrophys.\ J.\  {\bf 522}, 82 (1999); 
  A.~V.~Kravtsov,
  Adv.\ Astron.\  {\bf 2010}, 281913 (2010);  
   J.~Zavala, {\it et al.},
   Astrophys.\ J.\  {\bf 700}, 1779 (2009). 

\bibitem{deNaray:2011hy}
   R.~K.~de Naray and K.~Spekkens,
   Astrophys.\ J.\  {\bf 741}, L29 (2011); 
   M.~G.~Walker and J.~Penarrubia,
   Astrophys.\ J.\  {\bf 742}, 20 (2011).  

\bibitem{BoylanKolchin:2011de}
   M.~Boylan-Kolchin, J.~S.~Bullock and M.~Kaplinghat,
   Mon.\ Not.\ Roy.\ Astron.\ Soc.\  {\bf 415}, L40 (2011);  
   Mon.\ Not.\ Roy.\ Astron.\ Soc.\  {\bf 422}, 1203 (2012).
   
   \bibitem{Tulin:2012wi}
   S.~Tulin, H.~-B.~Yu and K.~M.~Zurek,
   Phys.\ Rev.\ Lett.\  {\bf 110}, 111301 (2013).

\bibitem{Tulin:2013teo}
   S.~Tulin, H.~-B.~Yu and K.~M.~Zurek,
   Phys.\ Rev.\ D {\bf 87}, 115007 (2013). 

\bibitem{Ko:2014nha}
  P.~Ko and Y.~Tang,
  arXiv:1402.6449 [hep-ph];
  arXiv:1404.0236 [hep-ph].

\bibitem{Kouvaris:2014uoa}
  C.~Kouvaris, I.~M.~Shoemaker and K.~Tuominen,
  Phys.\ Rev.\ D {\bf 91}, no. 4, 043519 (2015)
  [arXiv:1411.3730 [hep-ph]].

\bibitem{krstic:1999}
    P.~S.~Krsti\'{c} and D.~R.~Schultz, Phys.~Rev.~A {\bf 60}, 2118 (1999).

\bibitem{Drees:1992am}
  M.~Drees and M.~M.~Nojiri,
  Phys.\ Rev.\ D {\bf 47}, 376 (1993). 

\bibitem{susy-dm-review}
   G.~Jungman, M.~Kamionkowski and K.~Griest,
   Phys.\ Rept.\  {\bf 267}, 195 (1996).  

\bibitem{Dasgupta:2013zpn}
  B.~Dasgupta and J.~Kopp,
  Phys.\ Rev.\ Lett.\  {\bf 112}, no. 3, 031803 (2014)
  [arXiv:1310.6337 [hep-ph]].

\bibitem{Kusenko:2009up} 
  A.~Kusenko,
  Phys.\ Rept.\  {\bf 481}, 1 (2009)
  [arXiv:0906.2968 [hep-ph]];
  G.~M.~Fuller, C.~T.~Kishimoto and A.~Kusenko,
  arXiv:1110.6479 [astro-ph.CO].

\bibitem{atlas-diphoton}
 CMS Collaboration [CMS Collaboration],
  CMS-PAS-EXO-15-004.

\bibitem{cms-diphoton}
The ATLAS collaboration [ATLAS Collaboration],
ATLAS-CONF-2015-081.

\bibitem{diphoton-susy}
  C.~Petersson and R.~Torre,
  arXiv:1512.05333 [hep-ph];
  S.~V.~Demidov and D.~S.~Gorbunov,
  arXiv:1512.05723 [hep-ph];
  E.~Gabrielli, K.~Kannike, B.~Mele, M.~Raidal, C.~Spethmann and H.~Veermae,
  arXiv:1512.05961 [hep-ph];
  L.~M.~Carpenter, R.~Colburn and J.~Goodman,
  arXiv:1512.06107 [hep-ph];
  I.~Chakraborty and A.~Kundu,
  arXiv:1512.06508 [hep-ph];
  R.~Ding, L.~Huang, T.~Li and B.~Zhu,
  arXiv:1512.06560 [hep-ph];
  T.~F.~Feng, X.~Q.~Li, H.~B.~Zhang and S.~M.~Zhao,
  arXiv:1512.06696 [hep-ph];
  F.~Wang, L.~Wu, J.~M.~Yang and M.~Zhang,
  arXiv:1512.06715 [hep-ph];
  F.~Wang, W.~Wang, L.~Wu, J.~M.~Yang and M.~Zhang,
  arXiv:1512.08434 [hep-ph]
  B.~C.~Allanach, P.~S.~B.~Dev, S.~A.~Renner and K.~Sakurai,
  arXiv:1512.07645 [hep-ph];
  H.~Davoudiasl and C.~Zhang,
  arXiv:1512.07672 [hep-ph];
  N.~Craig, P.~Draper, C.~Kilic and S.~Thomas,
  arXiv:1512.07733 [hep-ph];
  J.~A.~Casas, J.~R.~Espinosa and J.~M.~Moreno,
  arXiv:1512.07895 [hep-ph];
  L.~J.~Hall, K.~Harigaya and Y.~Nomura,
  arXiv:1512.07904 [hep-ph];
  Y.~Jiang, Y.~Y.~Li and T.~Liu,
  arXiv:1512.09127 [hep-ph];
  E.~Ma,
  arXiv:1512.09159 [hep-ph].

\end{thebibliography}
\end{document}